\documentstyle[preprint,aps]{revtex}
\def \d {\delta}
\def \ds { \partial \raise.3ex \hbox {\kern -.55 em/}}
\def \ps { p \raise.3ex \hbox {\kern -.55 em/}}
\def \half {{1 \over 2}}
\def\overlay#1#2{\setbox0=\hbox{#1}\setbox1=\hbox to \wd0{\hss
#2\hss}#1\hskip -2\wd0\copy1}
\def\lsim{\mathrel{\rlap{\lower4pt\hbox{\hskip1pt$\sim$}}
    \raise1pt\hbox{$<$}}}         
\def\gsim{\mathrel{\rlap{\lower4pt\hbox{\hskip1pt$\sim$}}
    \raise1pt\hbox{$>$}}}         
\def\beq{\begin{equation}}
\def\eeq{\end{equation}}
\def\bea{\begin{eqnarray}}
\def\eea{\end{eqnarray}}
\def\nn{\nonumber}

\newcommand\bcdot{{\bf \cdot}}
\newcommand\bp{{\bf {p}}}
\newcommand\bq{{\bf {q}}}
\newcommand\bk{{\bf {k}}}
\newcommand\absp{{|\bf p|}}
\newcommand\absq{{|\bf q|}}

\newcommand{\bdelta}{\mbox{\boldmath$\delta$}}
\newcommand{\btau}{\mbox{\boldmath$\tau$}}
\newcommand{\bgamma}{\mbox{\boldmath{$\gamma$}}}
\begin{document}
\draft
\preprint{\vbox{Submited to {\it Physical Review \bf{C}}\hfill
IFUSP/P-1151\\}}
\title{Studying medium effects with the optimized $\bdelta$ expansion}
\author{G. Krein \\
Instituto de F\'{\i}sica Te\'orica, Universidade Estadual Paulista\\
Rua Pamplona 145, 01405-900 S\~ao Paulo-SP, Brazil}
\author{D.P. Menezes\\
Departamento de F\'{\i}sica, Universidade Federal de Santa Catarina\\
88.040-900 Florian\'opolis, S.C., Brazil}
\author{M. Nielsen\\
Instituto de F\'{\i}sica, Universidade de S\~ao Paulo \\
Caixa Postal 66318 - 5389-970 S\~ao Paulo, S.P., Brazil}
\author{M.B. Pinto\\
Laboratoire de Physique Math\'ematique, Universit\'e de Montpellier II \\
CNRS-URA 768, 34095 Montpellier Cedex 05, France}
\maketitle

\begin{abstract}
The possibility of using the optimized $\delta$ expansion for studying medium
effects on hadronic properties in quark or nuclear matter is investigated.
The $\delta$ expansion is employed to study density effects with two commonly
used models in hadron and nuclear physics, the Nambu--Jona-Lasinio model for
the dynamical chiral symmetry breaking and the Walecka model for the
equation of state of nuclear matter. The results obtained with the $\delta$
expansion are compared to those obtained with the traditional Hartree-Fock
approximation. Perspectives for using the $\delta$ expansion in other field
theoretic models in hadron and nuclear physics are discussed.
\end{abstract}
\newpage
\section{Introduction}
\label{sec:intro}

The study of possible modifications of hadron properties in the nuclear medium
is one of the central problems of contemporary nuclear physics. In principle,
these and related phenomena in nuclear physics are governed by the
fundamental theory of the strong interactions, quantum chromodynamics (QCD).
However, although QCD has been very successful in explaining a large
class of hadronic processes at high energy and large momentum transfer, typical
nuclear phenomena at lower energies cannot be derived from QCD with the
theoretical tools presently available. The difficulty of using QCD for
phenomena at the nuclear scale is related to the nonperturbative nature
of these. Due to the asymptotic freedom property of QCD, high energy
processes are calculable by perturbative techniques in the quark-gluon
coupling constant. On the other hand, since there are no reliable systematic
approximation schemes in field theory for performing nonperturbative
calculations, the construction of models is an important aspect of low energy
QCD. While there is considerable optimism that eventually one will be able to
solve QCD numerically on the lattice using supercomputers, the development of
analytical approximation methods are in urgent need to make contact with the
wealth of data on nonperturbative phenomena, presently available, or that will
be available when the new experimental facilities under construction start
operating. The $\delta$ expansion~\cite{original} is an example of a method
recently developed with the aim of studying nonperturbative phenomena in field
theory.

The idea of the $\delta$ expansion is to perturb the original theory by the
introduction of an artificial expansion parameter $\delta$, absent in the
original theory. The parameter $\delta$ is introduced in such a way that it
interpolates between the theory one wants to solve and another theory that
one knows how to solve. The $\delta$ expansion can be formulated in two
different forms, the logarithmic $\delta$ expansion~\cite{original} and the
linear $\delta$ expansion~\cite{linear}-\cite{su}. In this paper we consider
the linear form. Specifically, let $S$ be the action of the theory one wants to
solve, and $S_0$ the action of the soluble theory. Then, the interpolating
action $S(\delta)$ is defined as
\beq
S(\delta)=(1-\delta)S_0 + S \delta ,
\label{sdelta}
\eeq
so that $S(0)=S_0$ and $S(1)=S$. The next step involves the evaluation of
desired physical quantities as a perturbation series in powers of $\delta$,
which is then set equal to $1$ at the end. A crucial aspect of the method is
the recognition that $S_0$ involves arbitrary unknown (dimensionful and/or
dimensionless) parameters. If one were able to solve the new theory to all
orders in $\delta$, the unknown parameters would not play any role, since no
physical observable would depend on them. However, since we will be able to
solve the interpolated theory only to a finite order in $\delta$, there will
remain a residual dependence of the results on the parameters of $S_0$. These
arbitrary parameters must therefore be determined according to some criterion
and in fact there are many ways in which this can be
done~\cite{linear}-\cite{QM}. One physically appealing way to fix the unknown
parameters, which is the one adopted here, is the principle of minimal
sensitivity (PMS) introduced in Ref.~\cite{PMS}. This principle amounts to the
requirement that physical quantities should be at least locally independent of
the parameters. In the original applications of the method, the unknown
parameters were set to be equal to unity. The $\delta$ expansion, together
with the criterion of the PMS of physical observables, is known as the
{\em optimized} $\delta$ {\em expansion}. The convergence of the optimized
$\delta$ expansion has been proved in Ref.~\cite{ian}.

The different forms of the $\delta$ expansion have been successfully applied
to many different problems in quantum mechanics \cite{QM}, particle
theory~\cite{part,mar}, statistical physics~\cite{stat} and lattice
field theory~\cite{du,hugh}. Motivated by these successes, in this
paper we investigate the possibility of employing the linear $\delta$
expansion to study medium effects in hadron and nuclear physics using
typical field theoretic models. In the next section, we consider the use
of the $\delta$ expansion  in the study of density effects on the chiral
symmetry breaking in the Nambu--Jona-Lasinio (NJL) model. The PMS criterion
for the typical chiral quantities is applied following previous experience
with the method. We also investigate an alternative way of fixing the
parameters by applying the PMS to the energy density of the system. In
Section~\ref{sec:Walecka} we consider the Walecka model and study the effective
nucleon mass in nuclear matter. Both problems can be treated by obtaining the
propagators of the fields involved. Traditionally, the propagators are obtained
in the Hartree-Fock (HF) approximation. In the case of the NJL model this
approximation amounts to neglecting the corrections to the four-point vertex.
In the Walecka model, the HF approximation consists in obtaining self-
consistently the nucleon propagator with bare meson propagators, and neglecting
corrections to the meson-nucleon vertices. An additional approximation in the
Walecka model is the neglection of vacuum effects in the nucleon propagator.
With the purpose of comparing the $\delta$ expansion method with the
traditional HF approximation, we also neglect vertex corrections in both
models, and neglect the vacuum in the Walecka model. Conclusions and
discussions of the perspectives for future calculations appear in
Section~\ref{sec:concl}.

\section{Nambu--Jona-Lasinio model}
\label{sec:NJL}

In the limit of zero current quark masses, the two-flavor Lagrangian
density of the Nambu--Jona-Lasinio (NJL) model~\cite{njl} is given by
\beq
{\cal L}_{\rm NJL} = \bar q(i\ds)q+G\left[\left(\bar q q\right)^2-
\left(\bar q\gamma_5\btau q\right)^2\right]\;,
\label{NJLag}
\eeq
where the quark field operators $q=q(x)$ represent the doublet of $u$ and $d$
quarks. Since the model is non-renormalizable, one has a cutoff $\Lambda$ as
an extra parameter of the model, besides $G$.

According to Eq.~(\ref{sdelta}), one needs to introduce a Lagrangian
density ${\cal L}_0$ such that
\beq
{\cal L}_{\rm NJL}(\d)=(1-\delta){\cal L}_0+\delta {\cal L}_{\rm NJL}\;,
\label{NJLinterp}
\eeq
where the equations of motion derived from ${\cal L}_0$ can be solved as
exactly as possible. Since we are looking for solutions which break chiral
symmetry, the natural choice for ${\cal L}_0$ is
\beq
{\cal L}_0=\bar q (i\ds - \mu)q\;,
\label{L0}
\eeq
where $\mu$ is an arbitrary mass parameter introduced for dimensional reasons.
Therefore, the interpolated NJL Lagrangian density can be written as
\bea
{\cal L}_{\rm NJL}(\d)&=&(1-\delta)\left[\bar q (i\ds-\mu)q\right]+
\delta \left\{\bar q (i\ds)q+G\left[\left(\bar q q\right)^2-
\left(\bar q\gamma_5\btau q\right)^2\right]\right\} \nn \\
&=&\bar q (i\ds - \mu) q+\delta\left\{G\left[\left(\bar q q\right)^2-
\left(\bar q\gamma_5\btau q\right)^2\right]+\mu \bar q q\right\}\;.
\label{Linterp}
\eea

The evaluation of physical quantities is performed using perturbation theory
in the parameter $\delta$. The physical quantities of interest, whose values
characterize the chiral symmetry breaking, are the constituent quark mass
$M_q$, the quark condensate $<\bar q q>$ and the pion decay constant $f_\pi$.
The quark condensate for a given flavor is given by
\beq
<\bar q q>=-i \int \frac{d^4 p}{(2\pi)^4} {\rm Tr}\left[S(p)\right]\;,
\label{qbq}
\eeq
where the trace is over spinor and color indices. The pion decay
constant can be evaluated using the Pagels-Stokar~\cite{PS} formula:
\beq
if_{\pi}q_{\mu}\delta^{ab} = g_{\pi qq} \int \frac{d^4p}{(2\pi)^4}
{\rm Tr}
\left[\gamma_{\mu}\gamma_5 \tau^a S(q)\gamma_5 \frac{\tau^b}{2}
S(q+p)\right]\;,
\label{fpi}
\eeq
where the trace now is over spinor, flavor and color indices. The quark-pion
coupling is obtained from the Goldberger-Treiman relation.

In order to calculate these quantities one needs the quark propagator $S(p)$,
which can be obtained using Dyson's equation. Expressed in terms of the
self-energy $\Sigma(p)$ the quark propagator reads
\beq
S^{-1}(p)=S^{-1}_0(p)-\Sigma(p),
\label{Dyson}
\eeq
where $S^{-1}_0(p)$ is the inverse of the quark propagator corresponding to
${\cal L}_0$:
\beq
S^{-1}_0(p)={\not\!p}-\mu.
\label{freeS}
\eeq
$\Sigma(p)$ is calculated as a power series in $\delta$, and $S(p)$ is then
obtained by inverting Eq.~(\ref{Dyson}). Since the self-energy is calculated
perturbatively, this is still a perturbative scheme. It is the application of
the PMS to physical quantities which furnishes the nonperturbative character
to the $\delta$ expansion.

In zeroth order in $\delta$, one has
\beq
\Sigma^{(0)}(p)=0\;.
\label{Sigma0}
\eeq
Therefore, one can invert Eq.~(\ref{Dyson}) by using the well-known modified
Feynman $i\epsilon$ prescription for particles in a Fermi sea~\cite{hs}.
The in-medium quark propagator is then given by:
\beq
S^0(p)=\frac{{\not\!p}+\mu}{p^2-\mu^2+i\epsilon}+2\pi i
\frac{{\not\!p}+\mu}
{2E_0(p)}\delta\left(p^0-E_0(p)\right)\theta\left(P_F-\absp
\right)\;,
\label{MediumS0}
\eeq
where $E_0(p)=\left(\bp^2+\mu^2\right)^{\half}$, and $P_F$ is the Fermi
momentum, which relates to the density $\rho$ via $P_F=(\pi^2 \rho/2)
^{1/3}$.

At this zeroth order in $\delta$, no dynamical content from the model has been
used. The dynamics of the model starts to show up at order $\delta$. At first
order in $\delta$, the self-energy $\Sigma(p)$ is given by:
\bea
\Sigma^{(1)}(p)=&-&\delta \mu\nn\\
&+& 2i\delta G \int\frac{d^4q}{(2\pi)^4}\left\{{\rm Tr}\left[S_0(q)
\right]
- S_0(q)-\gamma_5 \tau^a {\rm Tr}\left[\tau^a S_0(q)\gamma_5\right] +
\gamma_5 \tau^a S_0(q) \tau^a \gamma_5 \right\}\;,
\label{SDENJL}
\eea
where a sum over the isospin index $a$ is implied.

Substituting Eq.~(\ref{MediumS0}) into this equation, we obtain for
$\Sigma^{(1)}$ the following expression:
\beq
\Sigma^{(1)}(p)=- \delta \mu+M_1-\gamma_0 \Sigma_0\;,
\label{sigma1}
\eeq
where
\beq
M_1=\delta{G\over \pi^2}\mu\left( N_cN_f+{1\over 2} \right )\left
\{ \Lambda\left(\Lambda^2+
\mu^2\right)^{\half}-P_F\left(P_F^2+\mu^2\right)^{\half} -\mu^2
\ln\left[{\Lambda + \left(\Lambda^2+\mu^2\right)^{\half} \over P_F +
\left(P_F^2+\mu^2\right)^{\half}}\right]\right\}\;,
\label{defM1}
\eeq
and
\beq
\Sigma_0= - 4 \delta G \int \frac{d^3q}{(2\pi)^3}\theta(P_F-\absq) .
\label{defSig0}
\eeq
Since the effect of $\Sigma_0$ is just to shift the chemical
potential~\cite{klev} one may write the constituent quark mass to $O(\delta)$
as
\beq
M_q= \mu - \delta \mu + M_1 .
\label{defMq1}
\eeq

Substituting Eq.~(\ref{MediumS0}) into Eq.~(\ref{qbq}) and Eq.~(\ref{fpi}), one
gets for the order parameter per flavor and for the pion decay constant the
following lowest order expressions:
\beq
<\bar q q>=-{2N_c\mu \over (2 \pi)^2}\left \{ \Lambda\left(\Lambda^2+
\mu^2\right)^{\half}-P_F\left(P_F^2+\mu^2\right)^{\half} -\mu^2
\ln\left[{\Lambda + \left(\Lambda^2+\mu^2\right)^{\half}
\over P_F +
\left(P_F^2+\mu^2\right)^{\half}}\right]\right\}\;,
\label{qbq0}
\eeq
and
\beq
f_{\pi}^2={N_cN_f\mu^2 \over (2 \pi)^2}\left \{ \ln \left[
{\Lambda + \left(\Lambda^2+\mu^2\right)^{\half} \over
P_F+(P_F^2+\mu^2)^{\half} } \right ] - \left( 1 + {\mu^2
\over \Lambda^2}\right )^{-\half} +\left(1+{\mu^2 \over P_F^2}
\right)^{-\half} \right\}\;,
\label{fpi0}
\eeq
where the lowest order Goldberger-Treiman relation,
$g_{\pi qq}=\mu/f_{\pi}$,
has been used.

The next step in the process is to fix $\mu$. In a previous work~\cite{mar}
this arbitrary parameter was determined by requiring $f_{\pi}$, which is
a fundamental quantity in the study of chiral symmetry breaking, to be
stationary with respect to variations in $\mu$. This is also a convenient
choice since, apart from having a well known empirical value, $f_{\pi}$ is the
only one of the studied quantities in Ref.~\cite{mar} which has a well defined
stationary point for finite values of $\mu$. To fix the noncovariant cutoff
$\Lambda$ one uses the empirical value $f_{\pi}$ at zero density. If one
applies the criterion of stationarity to $f_\pi$ in the vacuum, one obtains
(for $N_c=3$ and $N_f=2$)
\beq
\mu = 0.97 \times \Lambda\;.
\eeq
With the input $f_{\pi}= 93~~{\rm MeV}$, one finds $\Lambda=571$ MeV,
which implies $\mu=553$ MeV and $<\bar q q>= - (250~~{\rm MeV})^3$. Using
$G\Lambda^2=2.89$ as in Ref.~\cite{mar} and setting $\delta=1$ in
Eq.~(\ref{defMq1}) one finds the constituent quark mass to $O(\delta)$ to be
$M_q=574$ MeV. Figure 1 shows $\mu$ as a function of the density and has been
obtained by applying the PMS to $f_\pi$ for different values of $P_F$.
The results obtained for $M_q$ and $-<\bar q q>^{1/3}$, for different values
of $P_F$ are shown in Figure 2 (solid and dashed lines respectively). In Figure
3 the solid line shows the $P_F$ dependence of $f_{\pi}$. Contrary to what
happens at finite temperature~\cite{mar} we find that our result are
insensitive to whether $\mu(P_F)$ (as in Figure 1) or
$\mu(0) = 533~~{\rm MeV}$ is used.

A natural question which arises at this point is the uniqueness of the
of the value of $\mu$. If one were to use other physical quantities to fix
$\mu$, as for example the masses of the vector mesons, it is very  likely
that one would obtain a different value for $\mu$. In such a case, one
would have a different quark propagator for each observable. Of course this
would not be catastrophic if the spread of the values of $\mu$ determined with
different observables is not too large. In order to avoid such potential
uncertainties, we propose to fix $\mu$ by demanding that the energy density of
the system be stationary with respect to variations of $\mu$. The energy
density can always be written in terms of the propagators of the theory, and
then it is natural to demand stationarity of the energy with respect to the
unknown parameters of the propagators. Then, all physical observables are
determined from the same quark propagator.

 From the Lagrangian density, Eq.~(\ref{NJLag}), we have that the
energy-momentum
tensor is given by:
\bea
T^{\mu\nu}_{\rm NJL}&=&i\bar q \gamma^{\mu} \partial^\nu q
-g^{\mu\nu}{\cal L}_{\rm NJL}\nn\\
&=i&\bar q \gamma^{\mu} \partial^\nu q
-g^{\mu\nu}\left\{\bar q(i\ds)q + G\left[\left(\bar q q\right)^2-
\left(\bar q\gamma_5\btau q\right)^2\right]\right\}\;.
\label{Tmunu}
\eea
Note that we have not used the equation of motion for the quark field operator.
The energy density is the volume integral of the expectation value of
$T^{00}$ in the many-quark state. The expectation value of the field operators
can be evaluated using the usual Wick contraction technique. This leads
to
\bea
{\cal E_{\rm NJL}}&=& \frac{1}{V}\int d^3x\;<T^{00}>\nn\\
&=&-i\int\frac{d^4q}{(2\pi)^4}q^0{\rm Tr}\left[\gamma^0S(q)\right]+
i\int\frac{d^4q}{(2\pi)^4}{\rm Tr}\left[{\not\!q}S(q)\right]-
G\left\{-\left[\int\frac{d^4q}{(2\pi)^4}{\rm Tr}\left[S(q)\right]
\right]^2 \right. \nn\\
&&+\int\frac{d^4q}{(2\pi)^4}\frac{d^4k}{(2\pi)^4}{\rm Tr}
\left[S(q)S(k) \right] +\left[\int\frac{d^4q}{(2\pi)^4}{\rm Tr}
\left[\btau\gamma_5S(q)\right]\right]^2\nn\\
&&-\left. \int\frac{d^4q}{(2\pi)^4}\frac{d^4k}{(2\pi)^4}{\rm Tr}
\left[\gamma_5\tau^aS(q)\gamma_5\tau^a S(k)\right]\right\}\;.
\label{ENJL}
\eea

Substituting Eq.~(\ref{MediumS0}) into the expression above, we obtain:
\beq
{\cal E}^{(0)}_{\rm NJL}=-2N_cN_f\int_{P_F}^\Lambda
\frac{d^3q}{(2\pi)^3}\frac{\bq^2}{E_0(q)}-
2GN_cN_f(2N_cN_f+1) \left[\int_{P_F}^\Lambda
\frac{d^3q}{(2\pi)^3} \frac{\mu}{E_0(q)}
\right]^2\;.
\label{ENJLM0}
\eeq

The requirement that ${\cal E}$ be stationary with respect to variations in
$\mu$ leads to
\beq
\mu=4G\left(N_cN_f+\frac{1}{2}\right) \int_{P_F}^\Lambda \frac{d^3q}{(2\pi)^3}
\frac{\mu}{E_0(q)}\;.
\label{mustat}
\eeq
This is is the familiar Hartree-Fock gap equation of the model, where
$\mu$ has the interpretation of the dynamically generated mass.

Next, we consider the first order self-energy. By inverting Dyson's equation,
Eq.~(\ref{Dyson}), one obtains the quark propagator:
\beq
S^{(1)}(p)=\frac{{\not\!p}_1+M_1}{p^{2}_1-M^{2}_1+i\epsilon}+
2\pi i\frac{{\not\!p}_1+M_1}
{2E_1(p)}\delta\left(p^{0}_1-E_1(p)\right)\theta\left(P_F-\absp
\right)\;,
\label{MediumS1}
\eeq
where
\bea
&&p^{\mu}_1=(p^{0}_1, {\bp})=(p^0+\Sigma_0, \bp)\;,\\
&&E_1(p)=\left[\bp^2 + (M_1)^2 \right]^{\half}\;.
\label{aux1}
\eea
The superscript $(1)$ in $S^{(1)}$ indicates that the propagator has been
obtained with a self-energy calculated to first order in $\delta$. Note that
we are not expanding the propagator in powers of $\delta$. The process of
obtaining the propagator by inverting Dyson's equation with a self-energy
calculated in perturbation theory is known as the chain approximation.

Substituting Eq.~(\ref{MediumS1}) into Eq.~(\ref{ENJL}), we obtain:
\beq
{\cal E}^{(1)}_{\rm NJL}=-2N_cN_f\int_{P_F}^\Lambda
\frac{d^3q}{(2\pi)^3}\frac{\bq^2}{E_1(q)}-
2GN_cN_f(2N_cN_f+1) \left[\int_{P_F}^\Lambda
\frac{d^3q}{(2\pi)^3} \frac{M_1}
{E_1(q)} \right]^2\;,
\label{ENJLM1}
\eeq
where $M_1$ is given by Eq.~(\ref{defM1}) and $E_1(q)$ is defined in
Eq.~(\ref{aux1}).

Application of the PMS to ${\cal E}^{(1)}_{\rm NJL}$,
\beq
\frac{d{\cal E}^{(1)}_{\rm NJL}}{d\mu}=\frac{d{\cal E}^{(1)}_
{\rm NJL}}{dM_1}
\frac{dM_1}{d\mu}=0\;,
\eeq
leads to
\beq
M_1=4G\left(N_cN_f+\frac{1}{2}\right) \int_{P_F}^\Lambda
\frac{d^3q}{(2\pi)^3}
\frac{M_1}{E_1(q)}\;.
\label{M1stat}
\eeq
Again, we have obtained the familiar Hartree-Fock gap equation
for the dynamically generated mass.

If one proceeds to higher orders in $\delta$, in the scheme of neglecting
vertex corrections, the higher order quark propagator will always be of the
form of Eq.~(\ref{MediumS1}), with $M_1$ replaced by another constant, say $M$,
which is a function of $\mu$. However, because of the PMS condition on
${\cal E}$, $M$ at each order will always be given by the same value.
This value is the one that satisfies the usual gap equation:
\beq
M=4G\left(N_cN_f+\frac{1}{2}\right) \int_{P_F}^\Lambda
 \frac{d^3q}{(2\pi)^3}
\frac{M}{E(q)}\;,
\label{Mstat}
\eeq
where
\beq
E(q)=\left(\bq^2 + M^2\right)^{\half}\;.
\label{E}
\eeq
Therefore, the PMS condition on the energy density is equivalent
to the usual Hartree-Fock solution for the dynamically generated mass, in the
approximation of neglecting vertex corrections.

This result should be compared to the one presented in Ref.~\cite{su} where,
in the context of the effective potential, it was found that the $\delta$
expansion and the $1/N$ expansion are identical in the large $N$ limit.

In Figure 2 we compare the results obtained for the quark mass and
the quark condensate, when the two above described ways of applying
the PMS are used. We call PMS1 the results obtained by imposing $f_\pi$
to be stationary with respect to $\mu$, and PMS2 the results obtained
when the PMS is imposed to the energy density. The solid and dashed
lines give, respectively, $M_q$ and $-<\bar q q>^{1/3}$ obtained with PMS1,
and the dotted and dot-dashed lines give, respectively, $M_q$ and
$-<\bar q q>^{1/3}$ obtained with PMS2. In both cases we used the same set of
parameters: $\Lambda=571$ MeV and $G\Lambda^2=2.89$.

In Figure 3 we show the results obtained for $f_\pi$ with PMS1 (solid
line), PMS2 (dotted line), both with the parameters given above, and with
PMS2 (dashed line) with a new set of parameters: $\Lambda=653$ MeV and
$G\Lambda^2=1.98$. The last set of parameters was fixed by renormalizing
$f_\pi$ and $<\bar q q>$ at $P_F=0$ to their experimental values and requiring
$M_q$ to be roughly one-third of the nucleon mass. The curves for the
quark mass and condensate obtained by PMS2 with the renormalized parameters
are not shown because their behavior is analogous to $f_\pi$: they go to zero
at $P_F= 1.6$ fm$^{-1}$. At $P_F=0$ we have $M_q=314$ MeV. These are the usual
HF results and the renormalized parameters are the same as used in
Ref.~\cite{klev} for two flavors and a three momentum cutoff.

 From these figures we see that the results change appreciably when different
criteria are used. The main difference is related with the density
dependence of the quantities: while with PMS1 the quantities smoothly approach
zero at some critical density, they go to zero through a first order phase
transition with PMS2 (or Hartree-Fock)~\cite{klev}.

\section{Walecka model}
\label{sec:Walecka}

In this section we consider the Walecka model~\cite{wal} for nuclear matter.
The Lagrangian density of the model is given by
\beq
{\cal L}_{\rm W}=\bar \psi\left[\gamma_\mu(i\partial^{\mu}-g_\omega V^{\mu})-
(M-g_\sigma\phi)\right]\psi+\frac{1}{2}(\partial_{\mu}\phi\partial^{\mu}
\phi-m_\sigma^2 \phi^2)-\frac{1}{4}F_{\mu\nu}F^{\mu\nu}+\frac{1}{2}
m_\omega^2V_{\mu}V^{\mu}\;,
\label{LW}
\eeq
where $\psi$ represents the nucleon field operators, $\phi$ and $V_{\mu}$ are
respectively the field operators of the scalar and vector mesons, and
$F_{\mu\nu}=\partial_{\mu}V_{\nu}-\partial_{\nu}V_{\mu}$.

The energy-momentum tensor density corresponding to this Lagrangian density is
given by:
\beq
T^{\mu\nu}_{\rm W}=i\bar\psi\gamma^\mu \partial^\nu\psi+\partial^
\mu\phi
\partial^\nu\phi+\partial^\nu V_{\lambda}F^{\lambda\mu}-
g^{\mu\nu}{\cal L}_{\rm W}\;.
\label{TmunuW}
\eeq
Note that we have not used the nucleon equation of motion. Next, we eliminate
the meson field operators in favor of the nucleon field operators. The
Euler-Lagrange equations yield the meson field equations:
\bea
\left(\partial_\mu\partial^\mu+m_\sigma^2\right)\phi&=&g_\sigma
\bar\psi\psi\;,\label{eqsmes}
\\
\left(\partial_\mu\partial^\mu+m_\omega^2\right)V^\nu&=&
g_\omega\bar\psi\gamma^\nu\psi\;.
\label{eqsmev}
\eea
In obtaining the second equation above we have used baryon current
conservation, which implies that $\partial_\mu V^\mu=0$. These
equations can formally be integrated as:
\bea
\phi(x)&=&-g_\sigma\int d^4y\Delta_\sigma(x-y)\bar\psi(y)\psi(y)\;,
\label{phisol}
\\
V_{\mu}(x)&=&-g_\omega\int d^4y\Delta_\omega(x-y)\bar\psi(y)
\gamma_\mu\psi(y)\;,
\label{omegsol}
\eea
where $\Delta_i(x)$, $i=\sigma, \omega$, is given by:
\beq
\Delta_i(x)=\int \frac{d^4q}{(2\pi)^4}\frac{1}{q^2-m_i^2+i\epsilon}e^{-iqx} \;.
\label{mesprop}
\eeq
In Eq.~(\ref{omegsol}) above, because of baryon current conservation, we
have neglected the term proportional to $p^\mu p^\nu/m_\omega^2$ in the vector
meson propagator.

Using the expressions above for $\phi$ and $V^{\mu}$ in Eq.~(\ref{TmunuW}),
taking the expectation value of the resulting expression in the
many-nucleon state, and evaluating this with the help of Wick's contraction
technique, we obtain:
\beq
\langle T^{\mu\nu} \rangle=-i\int\frac{d^4q}{(2\pi)^4}\left\{{\rm Tr}
\left[\gamma^\mu q^\nu-g^{\mu\nu}({\not\!q}-M)\right]S(q)\right\}+
\langle T^{\mu\nu} \rangle_\sigma+\langle T^{\mu\nu} \rangle_\omega\;,
\eeq
with
\bea
\langle T^{\mu\nu} \rangle_\sigma &=&\frac{1}{2}\frac{g_\sigma^2}{m_\sigma^2}
\left[\int\frac{d^4q}{(2\pi)^4}{\rm Tr}S(q)\right]^2 g^{\mu\nu}-g_\sigma^2
\int\frac{d^4q}{(2\pi)^4} \frac{d^4k}{(2\pi)^4}{\rm Tr}\left[S(q+k)S(k)\right]
\Delta_\sigma(q^2) \nn\\
&\times&\left\{ \left[\frac{1}{2}(q^2-m_\sigma^2)\Delta_\sigma(q^2)-1\right]
g^{\mu\nu}-q^{\mu}q^{\nu}\Delta_\sigma(q^2)\right\},
\label{Tsigma}
\eea
and
\bea
\langle T^{\mu\nu} \rangle_\omega&=&
- \frac{1}{2}\frac{g_\omega^2}{m_\omega^2}\left[\int\frac{d^4q}{(2\pi)^4}
{\rm Tr}\gamma_\mu S(q)\right]\left[\int\frac{d^4q}{(2\pi)^4}{\rm Tr}
\gamma^\mu S(q)\right] g^{\mu\nu}\nn\\
&+& g_\omega^2\int\frac{d^4q}{(2\pi)^4}\frac{d^4k}{(2\pi)^4}
{\rm Tr}\left[\gamma_\lambda S(q+k)\gamma^\lambda S(k)\right]
\Delta_\omega(q^2)\nn\\
&\times&\left\{
\left[\frac{1}{2}(q^2-m_\omega^2)\Delta_\omega(q^2)-1\right]g^{\mu\nu}
- q^{\mu}q^{\nu}\Delta_\omega(q^2)\right\}\;.
\label{Tomega}
\eea

In the same way as in the NJL model, the nucleon propagator is obtained by
inverting Dyson's equation
\beq
S^{-1}(p)=S_0^{-1}(p)-\Sigma(p)\;,
\label{DysonN}
\eeq
where $S_0$ is the propagator corresponding to ${\cal L}_0$ in
Eq.~(\ref{Walinter}) below,
\beq
S_0^{-1}(p)={\not\!p}-M_0\;,
\eeq
with the self-energy $\Sigma(p)$ calculated as a perturbation expansion in
powers of $\delta$.

For infinite nuclear matter, because of the translational, rotational, parity
and time reversal invariances, $\Sigma(p)$ can be generally written in terms of
the unit matrix and the Dirac $\gamma_\mu$ matrices as follows~\cite{hs}:
\bea
\Sigma(p)&=&\Sigma^s(p)-\gamma_\mu \Sigma^\mu(p) \nn\\
&=&\Sigma^s(p^0,\absp)-\gamma^0\Sigma^0(p^0,\absp)+\bgamma\bcdot
\bp \Sigma^v(p^0, \absp) \;.
\label{genSig}
\eea
Defining the following auxiliary quantities \cite{hs}:
\bea
M^*(p)    &\equiv & M_0+\Sigma^s(p)\;,\nn\\
{\bp}^*   &\equiv & {\bp}\left[1+\Sigma^{v}(p)\right]\;,\nn\\
E^*(p)    &\equiv & \left[{\bp}^{*2}+M^{*2}(p)\right]^{\half}\;,\\
\label{auxW*}
p^{* \mu} & =     & p^\mu + \Sigma^\mu (p)=\left[p^0+\Sigma^0(p),
{\bp}^*\right]\;, \nn
\eea
we can invert Eq.~(\ref{DysonN}) and write the nucleon propagator in the
compact form:
\bea
S(p)&=&S_F(p)+S_D(p)\\
\label{SFSD}
S_F(p)&=&\left[\gamma^\mu p^*_\mu+M^*(p)\right]\frac{1}{p^{*\mu}p^*_{\mu}-
M^{*2}(p)+i\epsilon}\\
\label{SF}
S_D(p)&=&\left[\gamma^\mu p^*_\mu+M^*(p)\right]\frac{i\pi}{E^*(p)}
\delta\left(p^0-E(p)\right)\theta\left(P_F-\absp\right)\;,
\label{SD}
\eea
where $E(p)$ is the single-particle energy, which is the solution of the
transcendental equation:
\bea
E(p)&=&\left[E^*(p)-\Sigma^0(p)\right]_{p_0=E(p)}\nn\\
&=&\left\{\bp^2\left[1+\Sigma^v(\absp, E(p))\right]^2+\left[M+
\Sigma^s(\absp,E(p))\right]^2\right\}^{\half}-\Sigma^0(\absp, E(p))\;.
\label{selfE}
\eea
Note that we have assumed that the nucleon propagator has simple
poles with unit residue. Within the approximation scheme we are working in
this paper, this assumption is satisfied, as can be seen below.

Following the scheme of neglecting the Feynman part of the nucleon propagator,
Eq.~(\ref{SFSD}), we obtain for the energy density of nuclear matter the
following expression:
\bea
{\cal E}_{\rm W}&=&\frac{1}{V}\int d^3x T^{00}_{\rm W}-{\rm V.E.V.}
\nn\\
&=&\gamma\int_0^{P_F}\frac{d^3q}{(2\pi)^3}\frac{\bq\cdot\bq^*+
MM^*(q)}{E^*(q)}+{\cal E}^D_{\rm W}+{\cal E}^E_{\rm W}\;,
\label{EW}
\eea
where ${\rm V.E.V.}$ means the vacuum expectation value of $T^{00}_{\rm W}$,
and ${\cal E}^D_{\rm W}$ and ${\cal E}^E_{\rm W}$ are the direct and
exchange contributions, given by:
\bea
{\cal E}^D_{\rm W}&=& \left(\frac{1}{2}-1\right)\frac{g_\sigma^2}
{m_\sigma^2}\left[\gamma\int_0^{P_F}\frac{d^3q}{(2\pi)^3}\frac{M^*(q)}
{E^*(q)}\right]^2-\left(\frac{1}{2}-1\right)\frac{g_\omega^2}{m_\omega^2}
\left[\gamma\int_0^{P_F}\frac{d^3q}{(2\pi)^3}\right]^2\;,
\label{EWD}
\eea
\bea
{\cal E}^E_{\rm W}&=&\frac{1}{2}\gamma \int_0^{P_F}\frac{d^3q}
{(2\pi)^3E^*(q)}\int_0^{P_F}\frac{d^3k}{(2\pi)^3E^*(k)}\left\{g_\sigma^2
\Delta_\sigma(q-k) \right.\nn\\
&&\times\left[\left(\frac{1}{2}-1\right)-[E(q)-E(k)]^2
\Delta_\sigma(q-k)\right]\left[q^{*\mu}k^*_{\mu}+M^*(q)M^*(k)\right]\nn\\
&&+2g_\omega^2  \Delta_\omega(q-k)\left[\left(\frac{1}{2}-1\right)-
[E(q)-E(k)]^2\Delta_\omega(q-k)\right]\nn\\
&&\times\left.\left[q^{*\mu}k^*_{\mu}-2M^*(q)M^*(k)\right]
\right\}\;.
\label{EWE}
\eea
These expressions are very similar to the ones obtained in the Hartree-Fock
approximation. Differences are contained in the fermion kinetic energy,
the first term in Eq.~(\ref{EW}), and in the factors $\left(\frac{1}{2}
- 1\right)$ in Eqs. (\ref{EWD}) and (\ref{EWE}). These differences arise
because we are not using the nucleon field equation of motion.

To implement the $\delta$ expansion, we need to specify ${\cal L}_0$.
We choose
\beq
{\cal L}_0=\bar \psi\left(i\gamma_{\mu}\partial^{\mu}-M_0\right)\psi
+\frac{1}{2}(\partial_{\mu}\phi\partial^{\mu}\phi-m_\sigma^2\phi^2)-
\frac{1}{4}F_{\mu\nu}F^{\mu\nu}+\frac{1}{2}m_\omega^2V_{\mu}V^{\mu}\;,
\label{L0Wal}
\eeq
where
\beq
M_0 \equiv M+\mu\;.
\label{M0W}
\eeq
The interpolated Walecka model is then given by
\beq
{\cal L}_{\rm W}(\d)={\cal L}_0+\delta\left(-g_\omega\bar\psi\gamma_{\mu}
V^{\mu}\psi+g_\sigma \bar\psi\phi\psi+\mu\bar\psi\psi\right)\;.
\label{Walinter}
\eeq
Notice that the $\delta$ expansion technique could have also been applied
to the meson fields explicitly. However, we have chosen to eliminate the
meson fields in favor of the nucleon fields by means of Eqs.~(\ref{phisol}) and
(\ref{omegsol}), therefore meson effects enter via the nucleon fields.

Next we obtain the self-energy in perturbation theory, always neglecting vertex
corrections and the Feynman part of the nucleon propagator. In zeroth order in
$\delta$, the nucleon self-energy, corresponding to the interpolated Lagrangian
Eq.~(\ref{Walinter}), is obviously zero:
\beq
\Sigma^{(0)}=0\;.
\eeq
Therefore, the auxiliary quantities to be used in Eq.~(\ref{auxW*})
become:
\bea
M^*(p)    &\equiv & M_0 = M + \mu\;,\nn\\
{\bp}^*   &\equiv & {\bp}\;,\nn\\
E^*(p)    &\equiv & \left[{\bp}^{2}+M_0^2\right]^{\half}\;,\\
\label{auxW0}
p^{* \mu} & =     & p^\mu = (p^0,{\bp})\;.\nn
\eea
The single-particle energy is simply given by:
\beq
E(p)=E^*(p)=E_0(p)=\left[{\bp}^{2}+M_0^2\right]^{\half}\;.
\eeq
Using these in Eqs. (\ref{EWD}) and (\ref{EWE}), we obtain for the zeroth order
energy density the following expression:
\beq
{\cal E}^{(0)}_{\rm W}=\gamma\int_0^{P_F}\frac{d^3q}{(2\pi)^3}
\frac{\bq^2+MM_0}{E_0(q)}+{\cal E}_{\rm W}^{(0)D}+{\cal E}_
{\rm W}^
{(0)E}\;,
\label{EW0DE}
\eeq
with the direct and exchange contributions given by:
\bea
{\cal E}_{\rm W}^{(0)D}&=&\left(\frac{1}{2}-1\right)\frac{g_\sigma^2}
{m_\sigma^2}\left[\gamma\int_0^{P_F} \frac{d^3q}{(2\pi)^3}\frac{M_0}
{E_0(q)}\right]^2-\left(\frac{1}{2}-1\right)\frac{g_\omega^2}{m_\omega^2}
\left[\gamma\int_0^{P_F}\frac{d^3q}{(2\pi)^3}\right]^2\;,\\
\label{EW0D}
{\cal E}_{\rm W}^{(0)E}&=& \frac{1}{2} \gamma\int_0^{P_F}\frac{d^3q}
{(2\pi)^3E_0(q)}\int_0^{P_F}\frac{d^3k}{(2\pi)^3E_0(k)}\left\{g_\sigma^2
\Delta_\sigma([E_0(q)-E_0(k)]^2-(\bq-\bk)^2) \right.\nn\\
&\times&\left[\left(\frac{1}{2}-1\right)-[E_0(q)-E_0(k)]^2
\Delta_\sigma([E_0(q)-E_0(k)]^2-(\bq-\bk)^2)\right] \left[E_0(q)E_0(k)
\right.\nn\\
&-&\left.\bq\cdot\bk +M_0^2 \right]+2g_\omega^2
\Delta_\omega([E_0(q)-E_0(k)]^2-(\bq-\bk)^2)\left[\left(\frac{1}{2}-1\right)-
[E_0(q)-E_0(k)]^2\right.\nn\\
&\times&\left.\left.\Delta_\omega([E_0(q)-E_0(k)]^2-(\bq-\bk)^2)\right]
\left[E_0(q)E_0(k)-\bq\cdot\bk-2M_0^2\right]\right\}\;.
\label{EW0E}
\eea

Let us consider the direct term first. Application of the PMS to it:
\beq
\frac{d{\cal E}^{(0)}_{\rm W}}{d\mu}=\frac{d{\cal E}^{(0)}_{\rm W}}{dM_0}
\frac{dM_0}{d\mu}=\frac{d{\cal E}^{(0)}_{\rm W}}{dM_0}=0\;,
\eeq
yields the following self-consistency condition for $M_0$:
\beq
M_0=M-\frac{g_\sigma^2 }{m_\sigma^2}\gamma\int_0^{P_F}
\frac{d^3q}
{(2\pi)^3}
\frac{M_0}{E_0(q)}\;.
\label{M00}
\eeq
This is exactly the same self-consistency condition for the effective nucleon
mass obtained by means of the Hartree, or mean-field, approximation.

Now, application of the PMS to the full energy density, which includes both
direct and exchange contributions, leads to a nonlinear equation for $\mu$, or
equivalently for $M_0$, which is more complicated than the one of
Eq.~(\ref{M00}). We do not present the expression here because it is
rather lengthy and not very instructive. To investigate the size of the
exchange corrections we carry out two sets of comparisons. In Figure~4 we
compare the nucleon binding energy, obtained by using only the first and direct
terms in Eq.~(\ref{EW0DE}) (solid line) and coupling constants fixed by
fitting the binding energy and density of equilibrium nuclear matter, with
the full binding energy, keeping the same coupling constants (dotted line).
The value of the coupling constants are $g_s^2=91.64$ and $g_v^2=136.2$.
The masses used in all calculations are $M=939$ MeV, $m_v=783$ MeV and
$m_\sigma=550$ MeV. We find that the exchange corrections coincide with those
obtained in a relativistic Hartree-Fock calculation~\cite{wal,hs} which
we also show for comparison (long-dashed line). The dashed line shows the
results obtained for the total binding energy (including both direct and
exchange contributions) after renormalizing the model parameters to reproduce
the bulk saturation properties of nuclear matter: $g_s^2=83.11$ and
$g_v^2=108.05$. These coupling constants are the same used when renormalizing
the relativistic Hartree-Fock calculation of Ref.~\cite{hs}. Therefore, the
PMS condition on the zeroth order energy density of the Walecka model is also
equivalent to the usual Hartree-Fock solution.

In Figure~5 we show the results for $\mu$ as a function of the Fermi
momentum $P_F$ obtained with the application of the PMS to the zeroth order
energy density. The solid line corresponds to the first and direct terms only
and the
dashed one (almost unnoticeable) corresponds to the full energy density
with the
renormalized constants. In Figure~6 we compare the results for the effective
nucleon mass in nuclear matter as a function of $P_F$ obtained from $\mu$.
In both figures, it is clear that the results with the exchange terms
and renormalized constants coincide with the results obtained by using the
direct terms only.

We now consider the second-order contribution to the self-energy. The
self-energy to second-order in delta is is given by:
\bea
\Sigma^{(2)}(p)=&-&\mu \delta+i\frac{g_\sigma^2 \delta^2}{m_\sigma^2}
\int\frac{d^4q}{(2\pi)^4}{\rm Tr}\left[S^{(0)}(q)\right]-
i\frac{g_\omega^2 \delta^2}{m_\omega^2}\int\frac{d^4q}{(2\pi)^4}
\gamma_\mu {\rm Tr}\left[\gamma^{\mu}S^{(0)}(q)\right]
\nn\\
&+&ig_{\sigma}^2 \delta^2 \int\frac{d^4 q}{(2\pi)^4}S^{(0)}(q)
\Delta_{\sigma}(p-q)-ig_{\omega}^2 \delta^2 \int\frac{d^4 q}{(2\pi)^4}
\gamma_\mu S^{(0)}(q)\Delta_{\omega}(p-q)\gamma_{\mu}\;.
\label{SigmaW2}
\eea
where $\Delta_{\sigma}$ and $\Delta_{\omega}$ are given in Eq.~(\ref{mesprop}),
and again we have made use of the baryon current conservation. We evaluate
this expression neglecting the Feynman part of the nucleon propagator, the term
$S_F(p)$ given by Eq.~(\ref{SFSD}). Because of this, all integrals in
Eq.~(\ref{SigmaW2}) are finite and can easily be evaluated; there is no
need for renormalization. The first term in Eq.~(\ref{SigmaW2}) comes from the
first order contribution in $\delta$ and must be kept at second order. The
results are very similar to the ones obtained with the Hartree-Fock
approximation~\cite{hs}. Since there are subtle differences, we write them
explicitly below. Each component of the self-energy, $\Sigma^s$, $\Sigma^0$,
and $\Sigma^v$, can be decomposed in a direct and an exchange part. The direct
components are given by:
\bea
\Sigma^{s(2)}_D&=&-\delta \mu-\gamma \frac{g_\sigma^2 \delta^2}
{m_\sigma^2} \int_0^{P_F}
\frac{d^3q}{(2\pi)^3}\frac{M_0}{E_0(q)}\;,\\
\Sigma^{0(2)}_D&=&-\gamma\frac{g_\omega^2 \delta^2}
{m_\omega^2}\int_0^{P_F}\frac{d^3q}{(2\pi)^3}\;,\\
\Sigma^{v(2)}_D&=&0\;.
\label{sigmaD1}
\eea
Notice that they are independent of energy and momentum. The exchange terms are
given by:
\bea
\Sigma^{s(2)}_E(p)&=&\frac{1}{4\pi^2p}\int_0^{P_F}dq\;q\;\frac{M_0}{E_0(q)}
\left[\frac{1}{4}g_\sigma^2 \delta^2 \Theta_\sigma(p,q)-g_\omega^2 \delta^2
\Theta_\omega(p,q)\right]\;,\\
\Sigma^{0(2)}_E(p)&=&-\frac{1}{4\pi^2p}\int_0^{P_F}dq\;q\;\left[\frac{1}{4}
g_\sigma^2 \delta^2 \Theta_\sigma(p,q)+\frac{1}{2} {g_\omega^2 \delta^2}
\Theta_\omega(p,q)\right]\;,\\
\Sigma^{v(2)}_E(p)&=&-\frac{1}{4\pi^2p^2}\int_0^{P_F}dq\;q\;\frac{q}{E_0(q)}
\left[\frac{1}{2}g_\sigma^2 \delta^2 \Phi_\sigma(p,q)+g_\omega^2 \delta^2
\Phi_\omega(p,q)\right]\;,
\label{sigmaE1}
\eea
where the functions $\Theta_i(p,q), \Phi_i(p,q), i=\sigma, \omega$, are
defined by:
\bea
\Theta_i(p,q)&=&\ln \left| \frac{A_i(p,q)+2pq}{A_i(p,q)-2pq}\right|\;,\\
\Phi_i(p,q)&=&\frac{1}{4pq}A_i(p,q)\Theta_i(p,q)-1\;,
\label{ThetaPhi}
\eea
where
\beq
A_i(p,q)=\bp^2 + \bq^2 + m_i^2 -\left[E(p)-E_0(q)\right]^2\;.
\label{Apq}
\eeq

The auxiliary quantities to be substituted into Eq.~(\ref{auxW*})
are then given by:
\bea
M^*(p) &\equiv & M_0+\left[\Sigma_D^{s(2)} + \Sigma_E^{s(2)}(p)\right]\;,\nn\\
\bp^* &\equiv & {\bp}\left[1+\Sigma_E^{v(2)}(p)\right]\;,\nn\\
E^*(p)    &\equiv & \left[\bp^{*2}+M^{*2}\right]^{\half}\;,\\
\label{auxW2}
p^{* \mu} & = & p^\mu + \Sigma^\mu (p)=\left[p^0+\Sigma_D^{0(2)}+
\Sigma_E^{0(2)}(p), {\bp}^*\right]\;,\nn
\eea
and the single-particle energy is the solution of:
\bea
E(p)&=&\left[E^*(p)-\Sigma^{0(2)}(p)\right]_{p_0=E(p)}\nn\\
&=&\left\{\bp^2\left[1+\Sigma^{v(2)}(\absp, E(p))\right]^2+
\left[M+\Sigma^{s(2)}(\absp,E(p))\right]^2\right\}^{\half}-
\Sigma^{0(2)}(\absp, E(p))\;.
\label{selfE2}
\eea

We are in the position to calculate the energy density. Initially we consider
the direct terms only. The energy density is given by:
\beq
{\cal E}_W^{(2)D}=\gamma\int_0^{P_F}\frac{d^3q}{(2\pi)^3}\frac{\bq\cdot\bq^*+
MM^*}{E^*(q)} + \frac{g_\omega^2 }{2m_\omega}\left[\frac{2}
{3\pi^2}P_F^3\right]^2-\frac{g_\sigma^2 }{2m_\sigma^2}
\left[\gamma \int_0^{P_F}\frac{d^3q}{(2\pi)^3}
\frac{M^*}{E^*(q)}\right]^2\;.
\label{Hartree}
\eeq
Application of the PMS to this yields again the familiar Hartree
result, Eq. (\ref{M00}), with $M_0$ and $E_0(q)$ replaced respectively
by $M^*$ and $E^*(q)$. The exchange terms are shown in Eq.~(\ref{EWE}) and we
believe it is not necessary to rewrite them here. Also in this case, despite
numerical imprecisions, the exchange corrections coincide with the usual
Hartree-Fock solution, as can be seen in Figure~7. The behavior of $M^*$ as a
function of the Fermi momentum at this order does not show any noticeable
difference as compared with the zeroth order results. However, as can be seen
in Figure~8 the behavior of $\mu$ as a function of $P_F$ obtained with the
application of the PMS to the full second order energy density (dashed line)
is rather different from the one obtained when only the direct term is taken
into account (solid line). It is very interesting to notice that this different
behavior does not manifest itself either in the values of $M^*$ or of the
binding
energy. This is because the energy density $\cal E$ is a very flat function
of $\mu$, as can be seen in Figure~9, where the energy density is shown
for $P_F=1.19$ fm$^{-1}$. Recall that if one had an exact solution, the energy
density would be independent of $\mu$. The solid line is obtained without the
inclusion of the exchange term (the PMS solution in this case is given by
$\mu/M=-0.275$) and the dashed line gives the full second order density energy
(the PMS solution is $\mu/M=-0.35$). This stability in the value of the energy
density as a function of $\mu$ is very desirable and guarantees that even
big changes in the value of $\mu$ will not affect physical quantities, as the
binding energy for instance.

\section{Conclusions and Future Perspectives}
\label{sec:concl}

In this paper we have utilized the optimized $\delta$ expansion to study medium
effects in two commonly used models in hadron and nuclear physics: the
NJL model and the Walecka model. We have investigated an alternative way
of fixing the arbitrary parameters introduced by the $\delta$ expansion, by
applying the PMS to the energy density of the system.

The most important and concrete conclusion we can draw from this work
is that when applying the PMS to the energy density of the NJL model
we reproduce, at any order, the familiar Hartree-Fock solution for
the dynamically generated mass, in the approximation of neglecting
vertex corrections. In the case of Walecka model, we obtained results
quantitatively similar to
the ones of the usual Hartree-Fock approximation, although the analytical
expressions are not evidently equivalent. If one neglects the exchange term
in the energy density then clearly the mean-field solution is reproduced
at any order. It is also worth mentioning that, in the Walecka model, the
energy density is a very flat function of $\mu$ and this guarantees that
the PMS solution is indeed very stable.

On the basis of our results, we believe that the optimized $\delta$
expansion is a very robust nonperturbative approximation scheme. Compared with
the Hartree-Fock approximation, the $\delta$ expansion is very economical
because of its perturbative nature. Once the reliability of the scheme
has been established, one is ready to proceed to other interesting
applications. These include vertex and, obviously, vacuum effects. The
study of the vacuum in the Walecka model is an important issue since one
needs to know the limits of applicability of
such model to high densities and/or temperatures before quark and gluon
degrees of freedom have to be invoked. Of course, one has to face
renormalization problems when including the vacuum. Renormalization
in a Hartree-Fock scheme is very complicated~\cite{SerBiel} and one
expects that this will be facilitated within the $\delta$ expansion method.
\section{Acknowledgments}
This work was partially supported by CNPq and FAPESP.

%

%
\begin{figure}
\caption{$P_F$ dependence of $\mu$ in the NJL model obtained with PMS on
$f_\pi$. }
\label{mupF}
\end{figure}
\begin{figure}
\caption{Constituent quark mass (solid and dotted lines) and
$-<\bar q q>^{1/3}$ (dashed and dot-dashed lines)
as a function of $P_F$. The solid and dashed lines are the PMS1 solution
and the dotted and dot-dashed lines are the PMS2 solution.}
\end{figure}
\begin{figure}
\caption{$P_F$ dependence of $f_{\pi}$ for the NJL model. The solid and
dotted lines give respectively the PMS1 and PMS2 solutions
with the same parameters. The dashed line gives the PMS2 solution
with renormalized parameters.}
\end{figure}
\begin{figure}
\caption{$P_F$ dependence of the binding energy of the Walecka model at
zeroth order in $\delta$. The solid line represents the first and direct terms
of Eq. (59) only.
The dotted and long-dashed lines give the full binding energy and the
Hartree-Fock solution respectively, both determined with the same
coupling constants used in the solid line solution. Finally the dashed line
gives the full binding energy with the renormalized coupling constants.}
\end{figure}
\begin{figure}
\caption{$P_F$ dependence of $\mu$ for the Walecka model at
zeroth order in $\delta$. The dashed line represents
$\mu$ determined with the full self-energy, which is the sum of direct
and exchange terms. The solid line represents $\mu$ determined without
the exchange term.}
\label{mupFW}
\end{figure}
\begin{figure}
\caption{ Zeroth order nucleon effective mass $M_0$ as a function of $P_F$.
The solid curve is the result obtained without the exchange term
 and the dashed
curve is the result using the full energy density.}\label{FM00}
\end{figure}
\begin{figure}
\caption{$P_F$ dependence of the binding energy of the Walecka model at
second order in $\delta$. The solid, dashed, dotted and long-dashed lines
are the same as in fig.~4}
\end{figure}
\begin{figure}
\caption{$P_F$ dependence of $\mu$ for the Walecka model at
second order in $\delta$. The solid and dashed lines are the same as in
Figure~5.}
\label{mu2pFW}
\end{figure}
\begin{figure}
\caption{$ \mu$ dependence of the energy density for the Walecka model at
second order in $\delta$, calculated at $P_F=1.19$~~fm$^{-1}$. The solid line
gives the solution when the exchange term is not included. The dashed line
gives the full solution.}
\label{edmuW}
\end{figure}

\begin{references}
\bibitem{original} C.M. Bender, K.A. Milton, M. Moshe, S.S. Pinsky and
L.M. Simmons Jr., Phys., Rev. Lett. {\bf 58}, 2615 (1987).
\bibitem{linear} A. Okopi\'nska, Phys. Rev. D {\bf 35}, 1835 (1987).
\bibitem{du} A. Duncan and M. Moshe, Phys. Lett. B {\bf 215}, 352 (1988).
\bibitem{su}S.K. Gandhi, H.F. Jones and M.B. Pinto, Nucl. Phys. {\bf B359},
429 (1991).
\bibitem{PMS} P.M. Stevenson, Phys. Rev. D {\bf 23}, 2916 (1981).
\bibitem{QM}  C.M. Bender, K.A. Milton, M. Moshe, S.S. Pinsky and
L.M. Simmons
 Jr. Phys. Rev. D {\bf 37}, 1472 (1988); F. Cooper and P. Roy, Phys. Lett. A
{\bf 143}, 202 (1990).
\bibitem{ian} I. Buckley, A. Duncan and H.F. Jones, Phys. Rev. D {\bf 47},
2554 (1993); A. Duncan and H.F. Jones, {\it ibid.} D {\bf 47}, 2560 (1993);
C.M. Bender, A. Duncan and H.F. Jones, {\it ibid.} D {\bf 49}, 4219 (1994)
\bibitem{part} C.M. Bender and A. Rehban, Phys. Rev. D {\bf 41}, 3269 (1990);
C.M. Bender and K.A. Milton, Phys. Rev. D {\bf 38}, 1310 (1990); See also Ref.
 \cite{linear}; S.K. Gandhi and M.B. Pinto, Phys. Rev. D {\bf 49}, 4528 (1994).
\bibitem{mar} M.B. Pinto, Phys. Rev. D {\bf 50}, (1994), 7673.
\bibitem{stat} S.K. Gandhi and A.J. McKane, Nucl. Phys. {\bf B419}, 424
(1994).
\bibitem{hugh} I. Buckley and H.F. Jones, Phys. Rev. D {\bf 45}, 654 (1992);
J.O. Akeyo, H.F. Jones and C.S. Parker, Phys. Rev. D {\bf 51}, 1298 (1995).
\bibitem{PS} H. Pagels and S. Stokar, Phys. Rev. D {\bf 20}, 2947 (1979).
\bibitem{njl} Y. Nambu and G. Jona-Lasinio, Phys. Rev. {\bf 122}, 345 (1961).
\bibitem{klev} S. P. Klevansky, Rev. Mod. Phys. {\bf 64}, 649 (1992).
\bibitem{wal} J.D. Walecka, Ann. Phys. {\bf 83}, 491 (1974);
B.D. Serot and J.D. Walecka, Adv. Nucl. Phys. {\bf 16}, 1 (1985).
\bibitem{hs} C.J. Horowitz and B.D. Serot, Nucl. Phys. {\bf A399}, 529 (1983).
\bibitem{SerBiel} A.F. Bielajew and B.D. Serot, Ann. Phys. (NY) {\bf 156}, 215
(1984).
\end{references}
\end{document}